\documentclass[twocolumn]{aastex62}

\submitjournal{ApJ}

\shorttitle{ALMA unveils widespread molecular gas clumps in the Norma jellyfish galaxy}
\shortauthors{J\'achym et al.}
\begin{document}

\title{ALMA unveils widespread molecular gas clumps in the ram pressure stripped 
tail of the Norma jellyfish galaxy}

\correspondingauthor{Pavel J\'achym}
\email{jachym@ig.cas.cz}

\author{Pavel J\'achym}
\affil{Astronomical Institute, Czech Academy of Sciences, Bo\v cn\' i 
II 1401, 14100, Prague, Czech Republic}

\author{Jeffrey~D.~P. Kenney}
\affiliation{Department of Astronomy, Yale University, 260 Whitney 
Avenue, New Haven, CT 06511, USA}

\author{Ming Sun}
\affiliation{Department of Physics and Astronomy, University of Alabama 
in Huntsville, 301 Sparkman Drive, Huntsville, AL 35899, USA}

\author{Fran\c coise Combes}
\affiliation{Observatoire de Paris, LERMA, PSL, CNRS, Sorbonne Univ. 
UPMC, and College de France, F-75014, Paris, France}

\author{Luca Cortese}
\affiliation{International Centre for Radio Astronomy Research, The 
University of Western Australia, 35 Stirling Hwy, Crawley, WA 6009, 
Australia}

\author{Tom~C. Scott}
\affiliation{Institute of Astrophysics and Space Sciences (IA), Rua das 
Estrelas, P-4150-762 Porto, Portugal}

\author{Suresh Sivanandam}
\affiliation{Dunlap Institute for Astronomy and Astrophysics, 
University of Toronto, Rm 101, 50 St. George Street, Toronto, ON M5S 
3H4, Canada}

\author{Elias Brinks}
\affiliation{Centre for Astrophysics Research, University of 
Hertfordshire, College Lane, Hatfield AL10 9AB, UK}

\author{Elke Roediger}
\affiliation{Milne Centre for Astrophysics, Department of Physics \& 
Mathematics, University of Hull, Hull, HU6 7RX, UK}

\author{Jan Palou\v s}
\affiliation{Astronomical Institute, Czech Academy of Sciences, Bo\v cn\' i 
II 1401, 14100, Prague, Czech Republic}

\author{Michele Fumagalli}
\affiliation{Institute for Computational Cosmology and Centre for 
Extragalactic Astronomy, Durham University, South Road, Durham DH1 3LE, 
UK}



\begin{abstract}
We present the first high-resolution map of the cold molecular gas 
distribution, as traced by CO(2-1) emission with ALMA, in a prominent 
ram pressure stripped tail. The Norma cluster galaxy ESO~137-001 is 
undergoing a strong interaction with the surrounding intra-cluster 
medium and is one of the nearest jellyfish galaxies with a long 
multi-phase tail. We have mapped the full extent of the tail at 
$1\arcsec$ (350~pc) angular resolution and found a rich distribution of 
mostly compact CO regions extending to nearly 60~kpc in length and 
25~kpc in width. In total about $10^9~M_\odot$ of molecular gas was 
detected. The CO features are found predominantly at the heads of 
numerous small-scale ($\sim 1.5$~kpc) fireballs (i.e., star-forming 
clouds with linear streams of young stars extending toward the galaxy) 
but also of large-scale ($\sim 8$~kpc) super-fireballs, and 
double-sided fireballs that have additional diffuse ionized gas tails 
extending in the direction opposite to the stellar tails. The new data 
help to shed light on the origin of the molecular tail -- CO filaments 
oriented in the direction of the tail with only diffuse associated 
H$\alpha$ emission are likely young molecular features formed in situ, 
whereas other large CO features tilted with respect to the tail may 
have originated from the densest gas complexes that were pushed 
gradually away from the disk. The ALMA observations of ESO~137-001, 
together with observations from {\it HST}, {\it Chandra} and VLT/{\it 
MUSE}, offer the most complete view of a spectacular ram pressure 
stripped tail to date.
\end{abstract}

%ISM: structure
%ISM: molecules
%ISM: kinematics and dynamics
%(ISM:) evolution
%galaxies: interactions
%galaxies: star formation
%(galaxies:) intergalactic medium
%submillimeter: ISM


\section{Introduction} \label{sec:intro}

In recent years, numerous cluster galaxies with tails of ram pressure
stripped (RPS) gas have been discovered, many containing young stars
\citep[e.g.][]{cortese2006, sun2007, yoshida2008, smith2010, 
hester2010, yagi2013, ebeling2014, poggianti2017, gullieuszik2017, 
boselli2018}. These 
are dramatic examples of galaxy evolution driven by the cluster 
environment in which ram pressure of the intracluster medium (ICM) 
efficiently strips star-forming cool interstellar matter (ISM) from 
infalling galaxies \citep{gunn1972, koppen2018}. This causes star 
formation quenching of the main bodies of the galaxies 
\citep{vangorkom2004, koopmann2004} and their subsequent transformation 
towards early-types.

Another aspect of ram pressure stripping of galaxies is the 
production of new young stellar components in the stripped ISM which 
join galaxy halos or intracluster space \citep{sun2007, fumagalli2011, 
abramson2011, yagi2013, kenney2014, george2018, poggianti2019, 
cramer2019}. Young stars (ages $<10$~Myr) are found in the RPS tails at 
distances of up to several tens of kpc, indicating they have formed 
in-situ in the tails. Star formation in RPS tails provides a great 
opportunity to study the star formation process in a different 
environment from that in galaxy disks, something that is not 
sufficiently well understood in galaxy evolution. 

Tails of 'jellyfish' galaxies are multiphase, containing gas with a 
wide range of densities and temperatures. They have been detected in 
X-rays \citep[via bremstrahlung from hot ionized 
gas;][]{finoguenov2004, wang2004, machacek2005, sun2005, sun2006, 
sun2010}, H$\alpha$ \citep[both from compact HII regions photoionized 
by young stars, and from warm diffuse gas ionized by some other 
mechanism;][]{gavazzi2001, yoshida2004, yoshida2008, cortese2006, 
cortese2007, sun2007, yagi2007, yagi2010, fossati2012, fumagalli2014, 
boselli2016, poggianti2017}, and HI \citep[emitted by cooler neutral 
gas;][]{kenney2004, oosterloo2005, chung2007, chung2009, abramson2011, 
scott2010, scott2012, scott2018}.

Recently a cold molecular gas component has been detected in several 
tails \citep{jachym2014, verdugo2015, jachym2017, moretti2018}, in some 
cases forming an important fraction of the tail gas mass. In one 
sense, this is unsurprising given that stars are forming in the tails. 
On the other hand, explaining the presence of molecular clouds in ram 
pressure stripped tails is not trivial, since molecular clouds, tracing 
the densest parts of the ISM, resist direct stripping more than any 
other constituent of the ISM. Moreover, the survival of molecular 
clouds in the intra-cluster environment is poorly understood.

We would like to understand the formation and evolution of star-forming
molecular clouds in RPS tails. What are the roles of in-situ formation
versus direct stripping? Can stripped low density gas cool sufficiently
in the tails to produce dense gas and stars or does ram pressure need 
to be strong enough to directly strip dense gas? How do molecular gas 
complexes in tails (their mass distribution, lifetimes, SF 
efficiencies) compare to those in the disks of star-forming galaxies? 
How do star-forming molecular clouds evolve within the diverging gas 
flow of RPS tails? In order to make progress towards addressing these 
questions we need high resolution observations of molecular gas in the 
tails, a task that until now has not been possible.

We have obtained ALMA CO(2-1) observations of ESO~137-001, a textbook 
example of a galaxy that is undergoing stripping by ram pressure. This 
is the first time that a full map at sub-kpc resolution of the 
cold molecular gas distribution in a ram pressure stripped tail has been 
acquired.
To date, only off-center or off-disk regions of CO emission 
have been detected in maps of the RPS galaxies NGC~4848 and NGC~4522 and 
their close vicinity with the IRAM PdB and ALMA telescopes 
\citep{vollmer2001CO, lee2018}.

\begin{figure}
 \centering
 \includegraphics[width=0.45\textwidth]{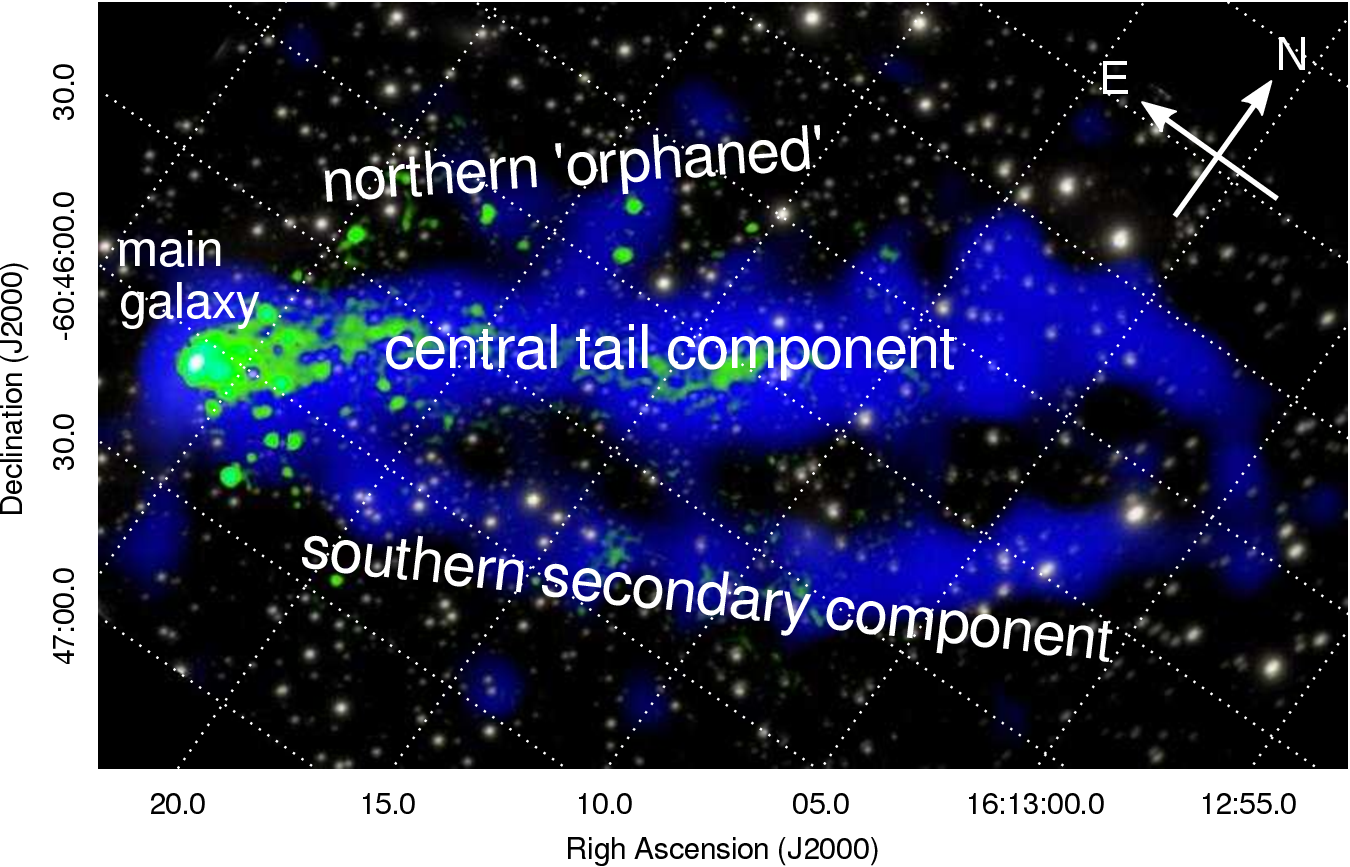}
\caption{
A composite X-ray {\it Chandra} (blue) / H$\alpha$ Southern Observatory 
for Astrophysical Research (SOAR) telescope (green) image of 
ESO~137-001 (Sun et al. 2010) illustrating the multi-component 
character of its ram pressure stripped tail -- the northern 'orphaned' 
regions, the central tail component, and the southern secondary 
component, are marked. Nearly all of the brightest optical sources are 
foreground stars. The image is rotated (note the N-E direction arrows) 
to match the orientation of the CO maps in Fig.~\ref{FigCO}. 
\label{FigESO}}
\end{figure}

\subsection{ESO~137-001}
ESO~137-001 galaxy in the Norma cluster (A3627, $M_{\rm dyn}\sim 
1\times 10^{15}~M_\odot$, $\sigma= 925$~km\,s$^{-1}$) is one of the 
nearest 'jellyfish galaxies' ($z=0.0163$), with a long X-ray, 
H$\alpha$, warm H$_2$, and CO tail, including young stars 
\citep{sun2006, sun2010, sun2007, sivanandam2010, fumagalli2014, 
jachym2014, fossati2016}. The presence of the one-sided gas tail, 
a strong HI-deficiency of the main galaxy \citep{vollmer2001}, and an 
undisturbed stellar disk are clear characteristics of a ram pressure 
stripping interaction. The galaxy is projected at only 
$\sim 14.5\arcmin= 270$~kpc from the cluster center (assuming $H_0= 
71$~km\,s$^{-1}$\,Mpc$^{-1}$, $\Omega_M= 0.27$, $1\arcsec\approx 
0.327$~kpc). The tail is observed in its full extent thanks to a 
favorable projection on the sky -- the galaxy's radial velocity 
relative to the cluster mean is only about $-200$~km\,s$^{-1}$, so its 
motion is mostly in the plane of the sky.

The multi-phase ESO~137-001 tail is also a multi-component system with 
a complex morphology -- see a composite X-ray/H$\alpha$ image in 
Fig.~\ref{FigESO}. The central tail component which is brightest and 
longest in X-ray and H$\alpha$ emission, connects to the main body of 
the galaxy. It is rather narrow with a width of $\sim 15\arcsec= 5$~kpc 
and a length of $\sim 210''= 70$~kpc, and a slightly bent shape. In 
addition there is also emission laterally beyond this central tail 
component. To the south there is a secondary X-ray and H$\alpha$ tail, 
which appears to originate outside the main body of the galaxy, above 
the stellar disk. To the north, there are some compact H$\alpha$ 
sources ('orphan HII regions' with no associated diffuse X-ray 
emission) and some patches of X-ray and H$\alpha$ emission. This part 
of the tail shows no gaseous emission connecting it to the stellar disk 
and there is also less diffuse (H$\alpha$ and X-ray) emission between 
the sources than in the southern tail component. Since RPS generally 
acts from the outside in (see Sec.~\ref{SecMorphology}), the different parts 
are likely at different evolutionary stages of stripping. 

\begin{figure*}
\centering
 \includegraphics[width=0.99\textwidth]{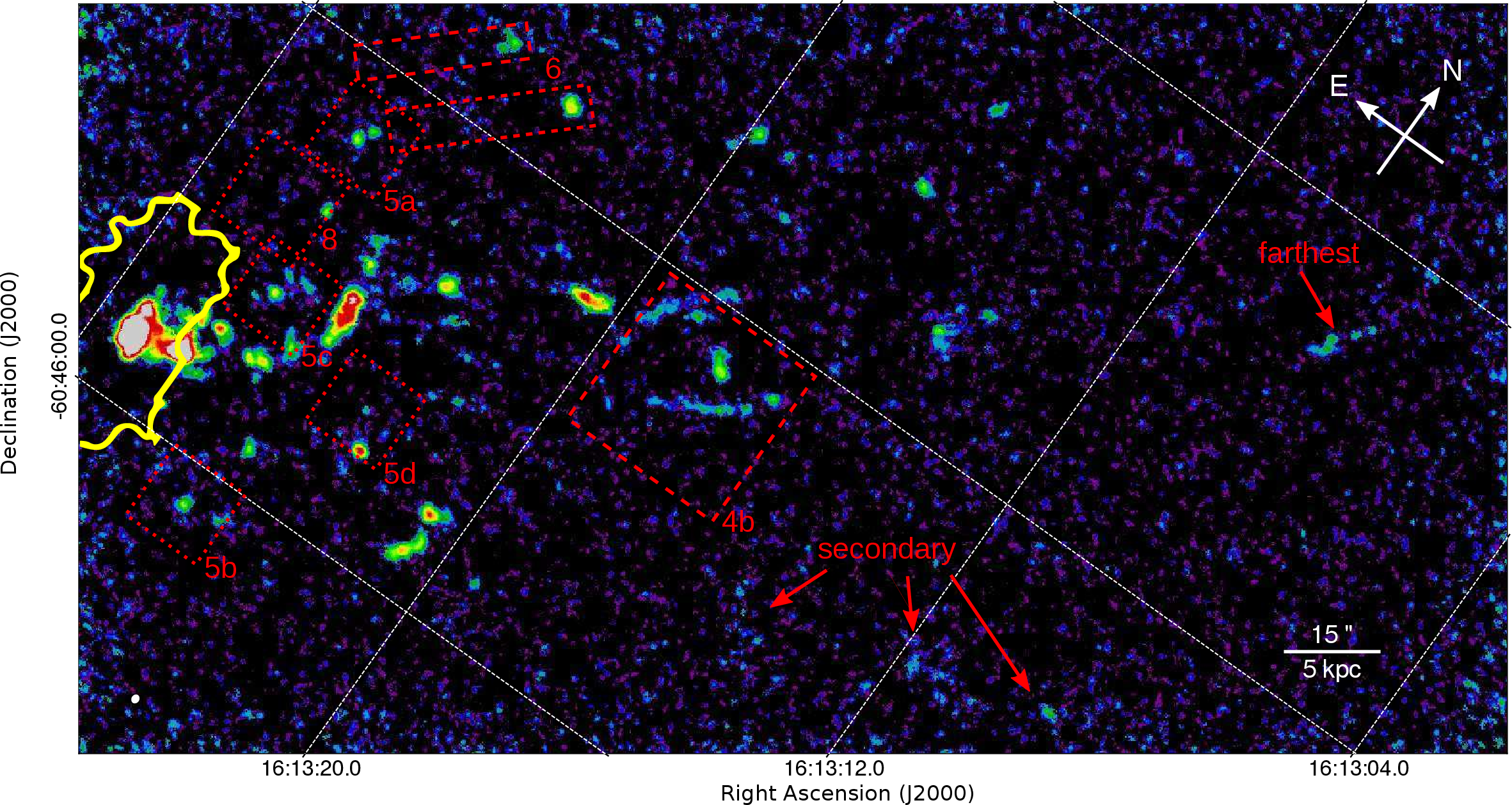}
 \includegraphics[width=0.99\textwidth]{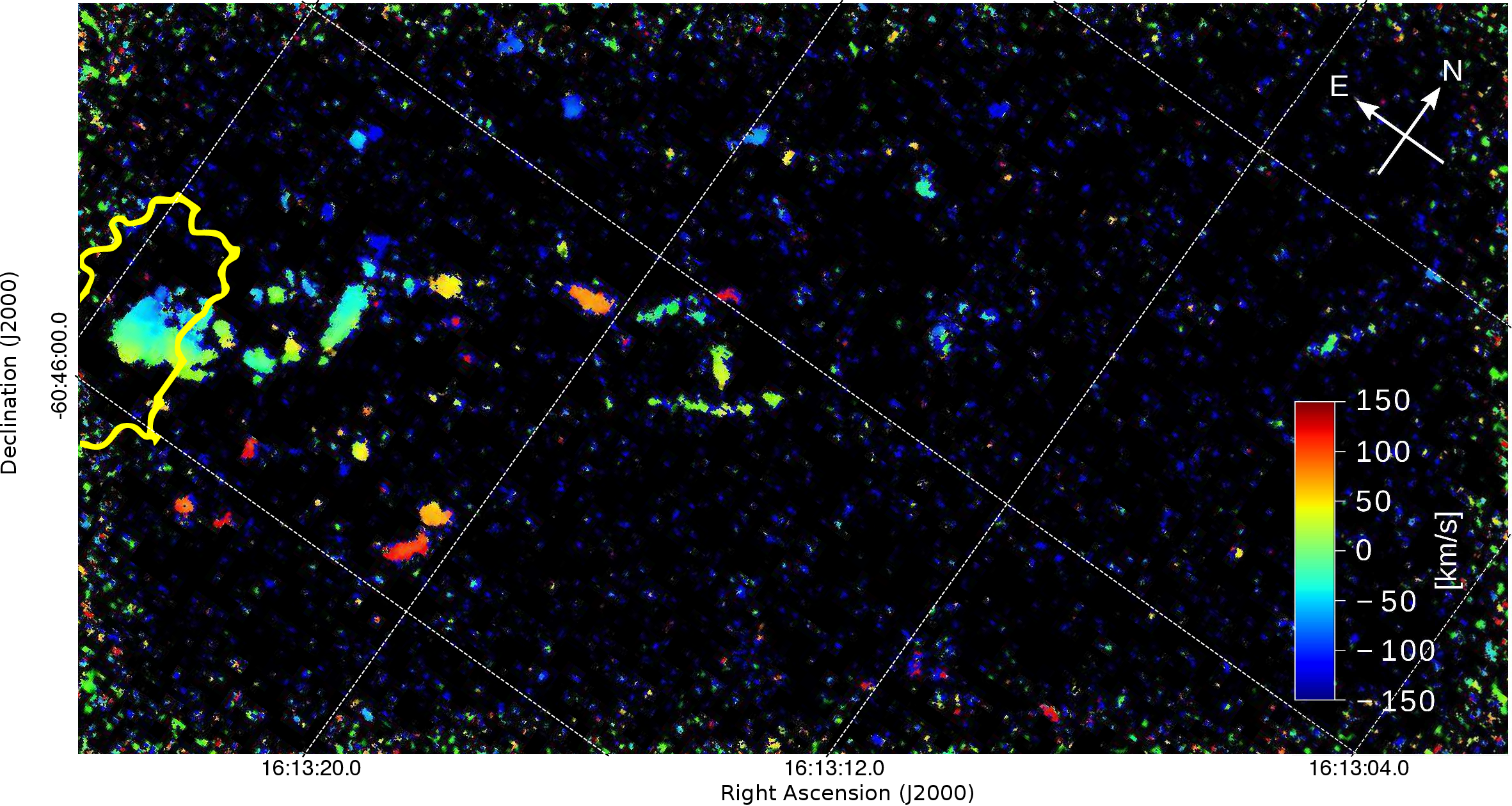}
 \caption{
Integrated brightness map (top) and velocity field map (bottom) of the 
CO(2-1) emission in ESO~137-001 and its tail. The image is rotated 
(note the N-E direction arrows). The synthesized beam size is $\sim 
1.4''\times 1.2''$, i.e. $\sim 460$~pc$\times 400$~pc (see the white 
dot in the lower left corner of the top panel). The outline of the 
optical disk is shown by drawing (yellow contour) the $r$-band isophote 
at 22~mag\ arcsec$^{-2}$ \citep{fossati2016}. Red dashed regions in the 
top panel mark the features displayed in Figs.~\ref{FigLarge} (panel b), 
\ref{FigFire} (panels a -- d), \ref{FigSuper} and \ref{FigDoubleObs}, 
and discussed in detail in Sec.~\ref{SecRes}. A couple of regions of 
interest are marked with arrows. CO clumps are detected in all three 
tail components -- the central tail, the northern 'orphaned' regions, 
as well as the secondary southern component, where some mostly 
low-S/N emission regions are found out to $\sim 30-40$~kpc from the 
galaxy (see arrows). The edges of the mosaics where noise strongly 
increases due to the applied correction for primary beam attenuation 
are truncated.
}\label{FigCO}
\end{figure*}

We used ALMA to observe all three tail components and nearly the full 
length of the (X-ray) tail - see the maps in Fig.~\ref{FigCO}. In the 
present paper we first describe the ALMA observations in 
Section~\ref{SecObs} and compare them with CO(2-1) single-dish 
observations. We present in Section~\ref{SecRes} maps of CO emission in 
ESO~137-001 and their comparison with ancillary data. 
We identify signatures of direct stripping and those indicating in-situ 
formation in Section~\ref{SecOri} and describe ``fireball'' 
features that are distinctive of ram pressure stripping in 
Section~\ref{SecDyn}. 
Our conclusions are given in Section~\ref{SecCon}.

\section{Observations}\label{SecObs}
The observations were carried out with ALMA from January to July, 2016 
(Cycle 3 project 2015.1.01520.S; PI: J\'achym). Using Band~6 in dual 
polarization mode, a spectral window was centered at the CO(2-1) line 
at a sky frequency of 227.955~GHz in frequency division mode (FDM) mode. 
The corresponding 
velocity resolution is $\sim 0.64$~km\,s$^{-1}$. Three spectral windows 
were set to observe the continuum, centered at sky frequencies of $\sim 
229.746$~GHz, 240.268~GHz, and 241.942~GHz. At $\sim 228$~GHz, the HPBW 
of the primary beam of the telescope was about $25.7\arcsec$. 

The main 12m array observations were completed in six sessions, with the 
number of antennas varying in the range $38-49$. The antennas were 
arranged in compact configurations with baselines in the range $15.1- 
469$~m. To cover the main galaxy and most of the spectacular tail, a 
$\sim 215\arcsec\times 123\arcsec$ rectangular mosaic of 131 pointings 
was set. An average on-source time per pointing was $\sim 1.9$~min. The 
median of the average system temperatures was 94~K. 

Additional 7m ACA array observations were obtained to increase the 
maximum recoverable spatial scale. In 15 observing sessions, $9-11$ 7m 
antennas were used, with minimum and maximum baselines of 8.9~m and 
48.9~m, respectively. The largest recoverable scale is $\sim 18\arcsec$ 
(calculated using the criterion of measuring 10\% of the total flux 
density of a uniform disk; see the {\it ALMA Cycle 3 Technical 
Handbook}, Table 7.1). An average $\sim 10$~min of on-source time was 
spent in each of 52 pointings covering the galaxy and its tail. The 
median of the average system temperatures was 75~K. 

Flux calibration was based on observations of Titan, J1617-5848 (12m 
and ACA) and J1427-4206 (ACA only). Bandpass calibration used J1427-4206 
(12m only), and J1924-2914 \& J1517-2422 (ACA only), and the phase 
calibrator was J1617-5848. 

The 12m array and ACA datasets were calibrated separately with the ALMA 
pipeline. Using the CASA software package 
\citep[v.4.5.3;][]{mcmullin2007}, the continuum was subtracted in the 
{\em uv-}plane using the task {\tt uvcontsub} first, followed by Briggs 
cleaning with a robust parameter of 0.5, and spectral smoothing to 
5~km\,s$^{-1}$. The resulting synthesized beam in the ACA and ALMA 
image datacubes is $\sim 7.9\arcsec\times 4.6\arcsec$ (PA of 
$88^\circ$) and $\sim 1.1\arcsec\times 1.0\arcsec$ (${\rm PA}= 
46^\circ$), respectively. The two datacubes were concatenated in a 
standard manner (after checking the ratio of the weights), using the 
CASA task {\tt concatenate}. The RMS noise levels in the line-free 
regions of the resulting datacube are $\sim 2.2$~mJy\,beam$^{-1}$ in 
5~km\,s$^{-1}$ channels. The final cube measures $2048\times 
2048$~pixels of $0.14\arcsec$ per pixel and 64 channels of 
5~km\,s$^{-1}$ width. The maps are corrected for primary beam 
attenuation.

In order to search for weak features, the maps presented in this paper 
were produced with a robust parameter of 2.0, thus close to natural 
weighting, which results in the lowest noise (highest signal-to-noise 
ratio). The resulting RMS level is $\sim 2$~mJy\,beam$^{-1}$, and the 
synthesized angular resolution $\sim 1.4\arcsec\times 1.2\arcsec$ 
(${\rm PA}= 58^\circ$). In order to create an integrated brightness 
CO(2-1) map, we explored different moment maps but due to the high 
dynamic range of the data many faint features stayed hidden in the 
noise. We therefore manually identified channels with maximum 
brightness in each pixel of the mosaic and integrated the spectrum only 
over a limited range of channels centered around the maximum channels 
\citep[using the {\tt astropy} python package;][]{astropy2013}. To 
determine the number of integrated channels we identified velocity 
widths of individual CO features and adapted the number of integrating 
channels accordingly. With this procedure we were able to reach 
better results than with the standard moment maps. The velocity field 
map was constructed using an analogous method.

The uncertainties in the CO flux are estimated to be about $5-20\%$. These 
include calibration uncertainties, CO flux errors arising from manually 
selecting the areas of the CO features, as well as the channels 
included in their integrated spectra.

\subsection{Compact-to-extended molecular ratio}\label{diffuse}
We use our previously obtained single-dish observations of ESO~137-001 
with APEX \citep{jachym2014} to make a direct comparison of the CO 
fluxes detected with APEX and ALMA in a number of common apertures. 
While in the main body of the galaxy the APEX and ALMA fluxes are 
consistent (flux ratio of $\sim 1.3$), in the tail regions the 
APEX-to-ALMA ratio increases to values of $3-4$. This suggests that 
while in the main body of the galaxy both telescopes reveal about the 
same amount of CO emitting gas, in the tail regions the ALMA 
observations (which included ACA) resolve out $\sim 67\%- 75\%$ of the 
CO emission. This is mainly due to missing total power observations and 
short-spacings leading to loss of extended, low surface brightness 
emission in the ALMA data. The RPS tail thus may be dominated by a 
diffuse, possibly warmer, subthermally excited molecular 
component \citep[e.g.][]{pety2013} distributed on spatial scales larger 
than $\sim 18\arcsec \approx 6$~kpc, which is the largest recoverable 
scale in our ALMA observations. Moreover, the ratio of the APEX-to-ALMA 
fluxes increases with distance along the tail, implying that the 
fraction due to an extended component becomes larger further out in the 
tail. 

We assume that for the compact features visible in the ALMA map and 
discussed further in this paper, the standard (Galactic) CO-to-H$_2$ 
conversion factor is applicable, while in the extended diffuse CO 
emitting component, it may be by a factor of a few lower 
\citep[e.g.,][]{bolatto2013}. The uncertainty in the value of the 
conversion factor does not affect our conclusions. Assuming that the 
extended CO emission is distributed smoothly along the tail, it is 
likely that locally the fraction of retrieved (compact) emission by 
ALMA is much larger, possibly near 90\%.  {\it Spitzer} IRS 
showed warm H$_2$ in the inner tail \citep{sivanandam2010}. What is 
the origin of the diffuse molecular component, and why it supposedly is 
stable against (gravitational) instabilities and thus does not form 
stars will be a matter for future studies.

\section{Results}\label{SecRes}
\subsection{Overall CO distribution and kinematics and relation to tail morphology}\label{SecCO}
\begin{figure*}
 \centering
 \includegraphics[width=0.95\textwidth]{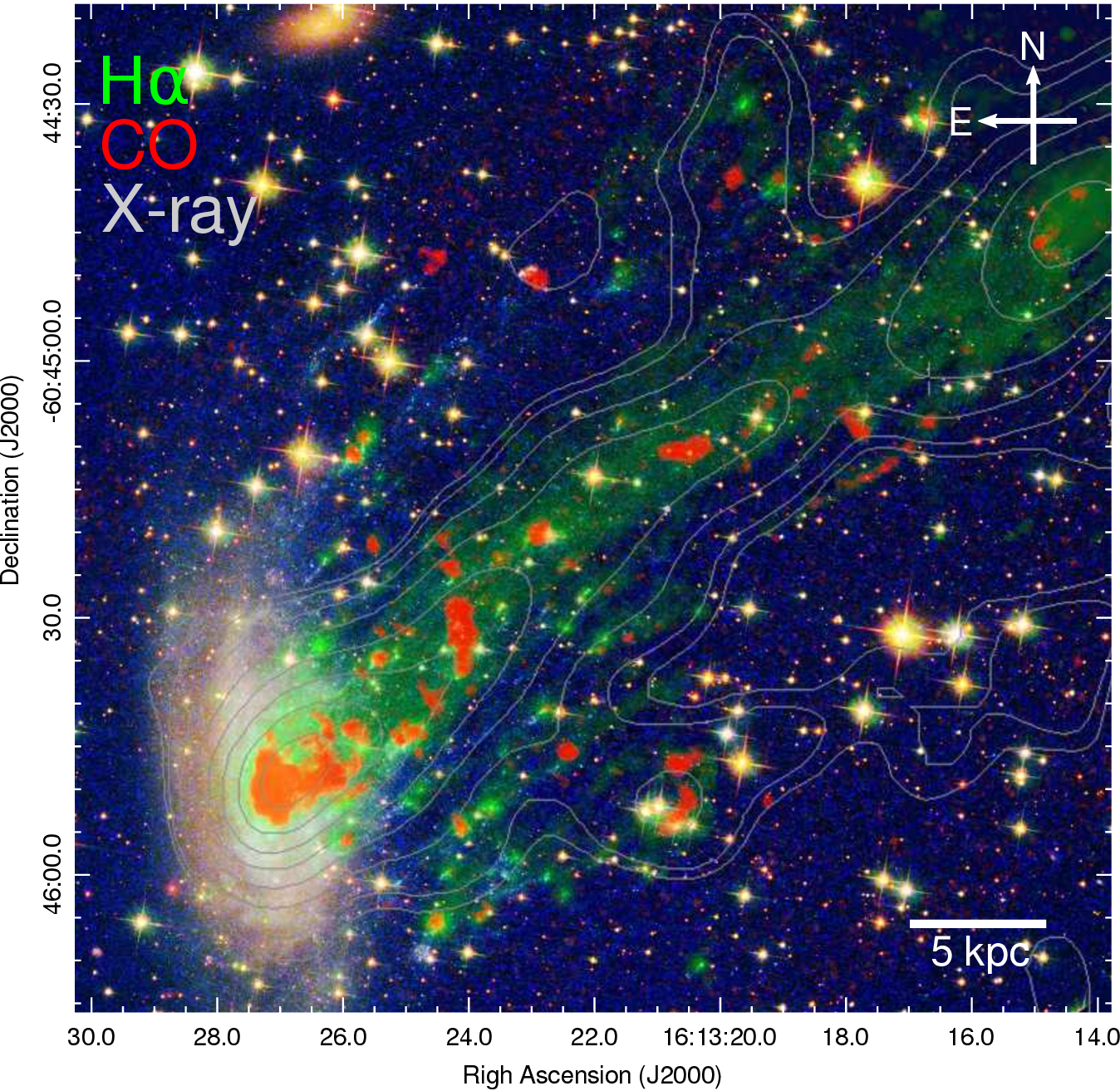}
 \caption{
Overlay of ALMA CO(2-1) emission (red) on MUSE H$\alpha$ emission 
(green) and the {\it HST} WFPC3 image of ESO~137-001 (blue: F275W; 
green: F475W; red: F814W) with {\it Chandra} X-ray contours on top 
\citep{sun2010}. North is up. Only the inner half of the (X-ray) tail 
is displayed. The upper left and lower right parts of the image were 
not covered by the MUSE observations \citep[see Fig.~2 
in][]{fumagalli2014}. The \textit{HST} image reveals a highly inclined 
spiral galaxy disk ($i\sim 66^\circ$) plus numerous complexes of blue, 
recently formed stars to the west of the disk, associated with parts of 
the ram pressure stripped gas tail.
}\label{FigHST}
\end{figure*}

In Fig.~\ref{FigCO} we show the total CO(2-1) intensity and velocity 
maps of ESO~137-001 and its tail, and in Fig.~\ref{FigHST} we show the 
CO intensity overlaid on \textit{HST} and H$\alpha$ images. In 
Fig.~\ref{FigCO} the maps are rotated (see the arrows indicating 
North and East directions). The RPS tail extends toward the right-hand 
side of the mosaic, in a NW direction on the sky. Besides the bright 
CO-emitting molecular gas in the main body of the galaxy, the map 
reveals a rich distribution of mostly compact CO emission regions 
detected in the tail out to nearly 60~kpc distance from the disk, and 
over a lateral extent of about 24~kpc. CO clumps are associated with 
all three tail components -- the northern 'orphaned' regions, the 
central tail, as well as the secondary southern tail. 

The CO emission in the main body of the galaxy is confined to the 
central region out to a radius of about $4-5''\sim 1.3-1.6$~kpc (S-N). 
This is the region of the galaxy disk where stripping is currently 
active. The stellar disk beyond this radius has already been stripped 
(except in the north where there is a CO clump at $\sim 7''$ from the 
disk center). The amount of CO 
flux in the disk area (within a $4''\times 10''$ area centered along 
the disk) is $\sim 43\pm 5$~Jy\,km\,s$^{-1}$. The integrated flux 
corresponds to an H$_2$ mass\footnote{We follow the standard relation 
for the CO(1-0) luminosity of \citet{solomon2005}, and assume the ratio 
of CO(2-1)-to-CO(1-0) of 0.8 \citep[e.g.,][]{leroy2009}, and a CO/H$_2$ 
conversion factor of $2\times 10^{20}$~cm$^{-2}$(K\,km\,s$^{-1})^{-1}$ 
that is standard under the Milky Way disk conditions 
\citep[e.g.,][]{bolatto2013}. A factor of 1.36 is included to account 
for helium.} of $\sim 6.9\times 10^8~M_\odot$, and an average face-on 
gas surface density within the central $r= 1.5$~kpc of about 
$150~M_\odot$\,pc$^{-2}$. Such a surface density is typical for the 
center of gas-rich spirals, and high enough to make the gas 
resistant to rapid ram pressure stripping.

Outside the main body, the brightest part of the tail is the central 
component with a total CO integrated flux of $57\pm 
10$~Jy\,km\,s$^{-1}$, which corresponds to $M({\rm H}_2)\sim 9.1\times 
10^8~M_\odot$. It extends to $\sim 58$~kpc from the disk. It is 
brightest in its innermost parts -- within a mere 2.5~kpc adjacent to 
the disk, about 40\% of the total CO flux is encompassed. By contrast, 
the lateral components of the tail contain much less molecular gas: the 
southern tail registers about $9\pm 2$~Jy\,km\,s$^{-1}$, which is $\sim 
1.4\times 10^8~M_\odot$ of H$_2$, whereas the northern 'orphaned' 
regions contain $\sim 6\pm 2$~Jy\,km\,s$^{-1}$, i.e. $M({\rm H}_2)\sim 
9.6\times 10^7~M_\odot$. Thus, about 80\% of the detected total tail CO 
flux is in the central component, while small fractions of $\sim 10\%$ 
are in the southern and northern regions. This indicates that the most 
imporant contribution to the molecular tail component comes from the 
relatively dense ISM most recently stripped from the central disk 
regions.

The total mass of the detected dense molecular gas in the tail is 
comparable to the total estimated X-ray gas mass of about $\sim 
10^9~M_\odot$, but larger than the upper limit estimate for neutral 
atomic gas of $\sim 5\times 10^8~M_\odot$ \citep{sun2006, vollmer2001, 
jachym2014}.

The velocity structure shown in the bottom panel of Fig.~\ref{FigCO} is 
dominated by the galaxy's rotation that is imprinted into the stripped 
gas: most of the northern emission regions are blue-shifted relative to 
the tail centroid while those in the south are mostly red-shifted. Due 
to the galaxy's orbital motion that is mostly in the plane of the sky, 
a velocity gradient along the tail is absent. A similar 
velocity structure was reported also from the early H$\alpha$ 
observations \citep{sun2010} and later confirmed from the detailed MUSE 
H$\alpha$ map \citep{fumagalli2014}. The CO velocity dispersion is 
typically $\sim 5- 15$~km\,s$^{-1}$ in most clumps but larger ($\sim 
15- 25$~km\,s$^{-1}$) in the galaxy and the brightest regions of the 
inner tail.

\subsection{Compact molecular component of the tail}
The spatial resolution of the present ALMA observations of $\sim 
350$~pc does not allow us to resolve individual GMCs that in galaxies 
have typical sizes of the order of $\sim 1- 100$~pc. Besides some 
unresolved clumps, most of the detected CO emission regions in the tail 
of ESO~137-001 are larger structures, some with a clear substructure. 
Typical CO clumps have masses characteristic of GMCs or giant molecular 
associations (GMAs) of $\sim 10^6- 10^7~M_\odot$, but their sizes (and 
also velocity dispersion) are larger ($\sim 500- 800$~pc). This 
suggests that the clumps are not gravitationally bound and will likely 
disperse with time. For the selection of the CO clumps described in the 
present work, we can estimate their virial parameter $\alpha$. 
Following \citet{bertoldi1992},
\begin{equation}
%\alpha\equiv M_{\rm vir}/M= 5\sigma^2 R/(G\,M),
\alpha= 5\sigma^2 R/(G\,M),
\end{equation}
where $\sigma$, $R$, and $M$ are a clump's velocity dispersion, radius 
and mass, respectively (see the CO spectra in Fig.~\ref{FigFire}). The 
parameter $\alpha$ spans a range of $\sim 2- 12$ which indicates 
subcritical structures \citep[non-magnetized clouds are expected to 
have critical virial parameters $\sim 2$; e.g.,][]{kauffmann2013}. This 
supports the above mentioned suggestion that 
molecular clumps will dissolve with time, however the uncertainty of 
these estimates may be large. A detailed study of the molecular clump 
population detected in the ALMA data in the tail and its relationship 
to the star-forming regions will be the subject of a future paper. Also 
in the narrow RPS tail of the Coma galaxy D100, stellar sources were 
measured to be much larger than single gravitationally bound star 
clusters, indicating they are likely to be unbound and disperse 
\citep{cramer2019}.

\subsection{Multi-phase view of the RPS tail}
In Fig.~\ref{FigHST}, the CO(2-1) intensity map (red) is overlaid with 
H$\alpha$ (green) and \textit{HST} UV-visible-nearIR composite images, 
covering about the inner half of the length of the (X-ray) tail. 
Contours of X-ray emission are added into the overlay \citep{sun2010}. 
The image indicates various levels of correlation between the CO, 
H$\alpha$ and young stars. It also clearly shows that the narrow 
central tail is connected to the CO-emitting molecular gas still in the 
central parts of the main body of the galaxy. The CO clumps detected 
outside the main body of the galaxy are associated with all tail 
components, but in comparison to the more extended H$\alpha$ 
distribution, the CO emission comes from more compact regions. Given 
that the CO emission traces a relatively dense gas phase (see 
Sec.~\ref{diffuse}), the ALMA CO observations reveal for the first time 
the dense "skeleton" of the multi-phase gaseous tail, as well as the 
component that forms the stars that are visible in the {\it HST} image. 
The overlay image in Fig.~\ref{FigHST} offers the most complete view of 
the spectacular RPS tail up to now, with the CO, H$\alpha$, and X-ray 
emitting gas components, as well as the regions of young star 
formation.

\subsection{Molecular gas tail morphology and RPS history}\label{SecMorphology}
The overall molecular gas distribution in the tail roughly 
follows the distribution of the other gas phases and reflects the 
stripping history of the disk. The lateral extent of CO in the tail is 
much broader than the lateral extent of CO in the disk. This is 
because RPS generally acts from the outside in, and outer disk gas is 
generally stripped earlier and forms the broadest parts of the tail. 
The broadest parts of the tail (the northern and southern components) 
originated from outer disk regions that were fully stripped some time 
ago. The kinematics support this with red- and blue-shifted velocities 
in the parts of the tail furthest from the tail centroid 
(Fig.~\ref{FigCO}). 

The complex morphology of the multi-component tail of ESO~137-001 is 
not consistent with the simple outside-in stripping. 
The continuous central long component of the tail that is also bright 
in CO and that connects to the CO in the main body of the galaxy, 
extends much further out than the innermost northern 
tail component, and it is also spatially separated from both the 
northern and southern components. This indicates that the central few 
kpc of the galaxy has been experiencing stripping at the same time as 
the outer disk. 
In a galaxy with ISM substructure, the ICM wind may penetrate through 
low density parts of the disk, and multiple radii may be stripped 
simultaneously \citep[e.g.,][]{quilis2000}. This is what we think is 
likely happening in ESO~137-001.

The timescale of gas stripping depends on the galaxy's orbit in 
the cluster, as well as on the gas column density. Semi-analytic 
modeling indicates that ESO~137-001 is currently only $\sim 100$~Myr 
before pericenter along a highly radial orbit with a large orbital 
speed $>3000$~km\,s$^{-1}$ and the corresponding peaked ram pressure 
time profile having a FWHM $\sim 200$~Myr \citep{jachym2014}. For the range 
of column densities corresponding to the mean values of the typical CO 
clumps detected in the tail, we can calculate the distances along the 
tail which stripped disk gas parcels have reached and their 
corresponding travel times: from e.g. 2~kpc radius in the disk, a gas 
clump with the column density of 10 (20 or 30)~$M_\odot$\,pc$^{-2}$ 
was pushed to 80 (20 or 7)~kpc distance along the tail in the last 200 
(90 or 50)~Myr. From outer disk radii, stripping has started much 
earlier along the orbit and the stripped gas reaches much larger 
distances. These calculations neglect all hydrodynamic effects as well 
as the internal structure of gas clouds. It is interesting to compare 
the above estimates with the travel times if a constant value of the 
current RP ($\sim 1.5\times 10^{-10}$~dyne\,cm$^{-2}$) is applied 
instead of the time-evolving RP and the restoring force of the galaxy 
is neglected: 83 (58 or 42)~Myr, respectively.

The new observations also show the asymmetry between the northern and 
southern tail components. There are prominent isolated CO clumps in 
the northern tail occurring to nearly 40~kpc from the main galaxy, 
whereas in the southern tail clumps are found only out to $\sim 20$~kpc 
(with some marginal detections at larger distances). The asymmetry of 
the tail is likely caused by the effects of disk inclination with 
respect to the ram pressure direction. The interaction with the ICM 
wind is not face-on but tilted\footnote{The wind angle, i.e., the angle 
between the disk plane and the direction of the wind, has a value of 
$\sim 47^\circ$, following from the disk position angle ($\sim 
9^\circ$), the disk inclination angle ($\sim 66^\circ$, HyperLEDA), the 
position angle of the tail ($\sim 316^\circ$) and an assumption that 
the tail follows the past orbital path. The wind angle is thus a 
complementary angle between the wind vector $\vec{w}=(\sin{44^\circ}, 
\cos{44^\circ}, 0)$ and the normal vector to the disk plane 
$\vec{n}=(\cos{9^\circ}, \sin{9^\circ}, \cos{66^\circ})$.}
by $\sim 47^\circ$ which means that the rotation of the galaxy affects 
the trajectories of the stripped gas in an asymmetric manner, depending 
on whether the gas rotates with or against the ICM wind.

Thus different parts of the tail are at different evolutionary stages 
of stripping: the central component is still being fed by the galaxy, 
while the outer, northern and southern components represent later 
evolutionary stages, since these parts of the tail are no longer being 
fed from the galaxy. The northern tail component appears to be at later 
evolutionary stage than the southern component, as it has less diffuse 
H$\alpha$ and X-ray emission with isolated ('orphaned') dense CO 
and H$\alpha$ clumps apparently left behind the more diffuse gas. 
Despite their different ages, all three tail components are found to 
host molecular gas and also young stars.

\section{Origin of large molecular structures detected in the tail}\label{SecOri}
In the following section we focus on several distinct 
features that may prompt a new perspective on the origin and 
evolution of dense gas in the stripped ISM.
Two physical scenarios are usually considered, either the molecular 
gas has formed in situ out of stripped atomic gas ({\it in-situ 
formation}), or some pre-existing molecular gas contributed to it ({\it 
direct stripping}).
The central tail component is in the earliest evolutionary stage, in 
the sense that gas is still being fed to it from the main galaxy. We 
therefore might expect to find within it molecular gas concentrations 
at relatively early evoelutionary stages, including some that have 
recently formed. We describe two distinct CO complexes in the central 
tail, one of which may have formed from a large gas structure 
originating in the galaxy disk, and another which may have formed in 
situ.

\subsection{Large feature tilted $\sim 60^\circ$ from the tail axis}\label{Sec60}
\begin{figure*}
 \centering
 \includegraphics[height=0.24\textwidth]{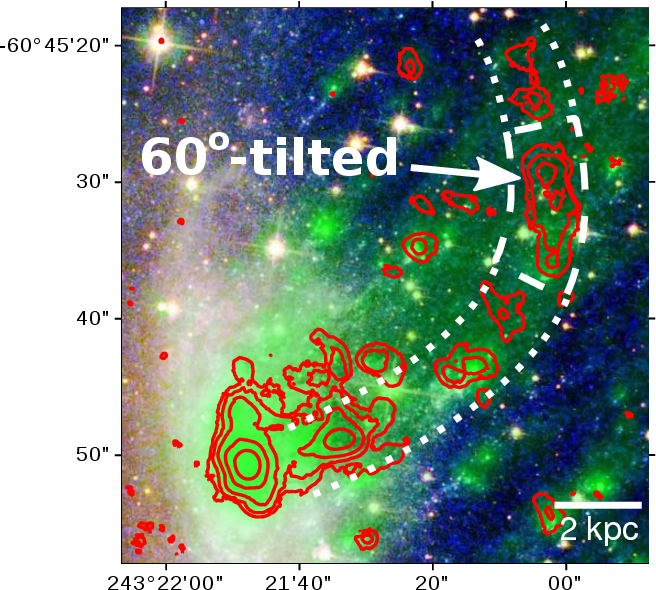}
 \includegraphics[width=0.23\textwidth]{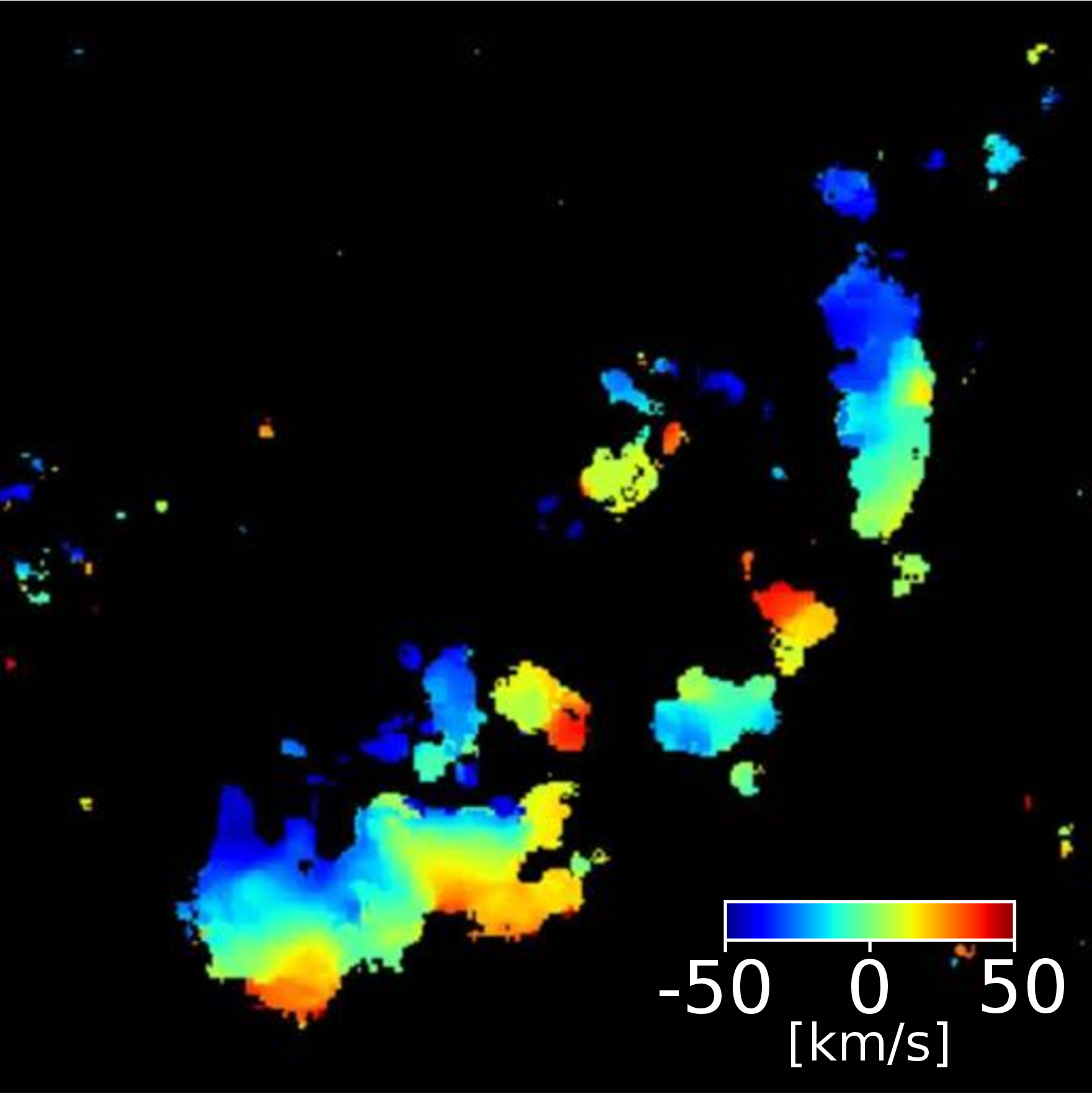}
 \includegraphics[height=0.23\textwidth]{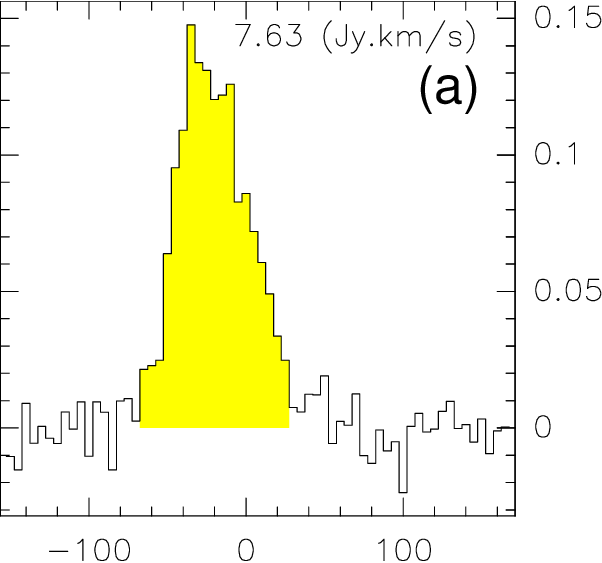}\\
 \includegraphics[height=0.24\textwidth]{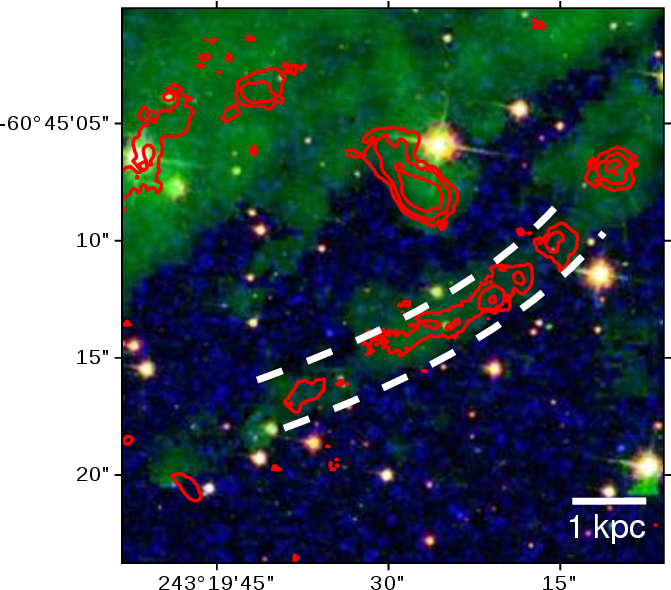}
 \includegraphics[width=0.23\textwidth]{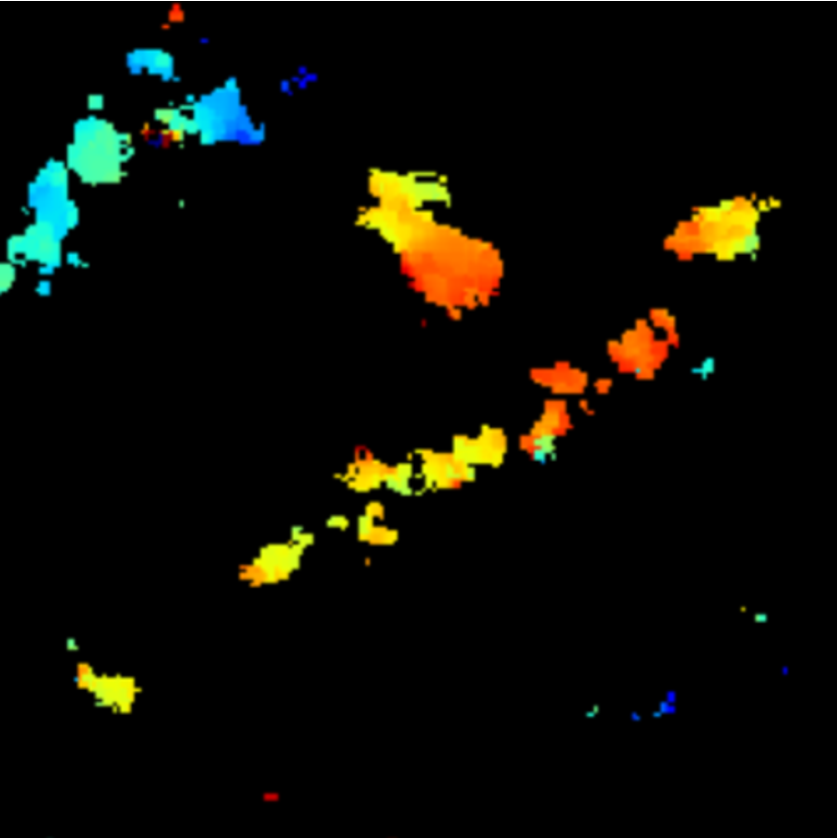}
 \includegraphics[height=0.23\textwidth]{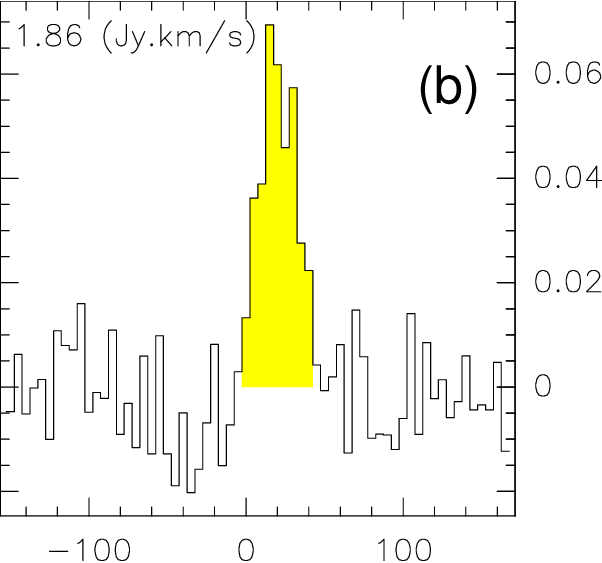}
 \caption{
Large CO features found in the central tail component -- the feature 
tilted $60^\circ$ from the tail axis (top row) and the filamentary 
structure (bottom row) -- shown in a composite {\textit HST}-MUSE image 
with CO contours overlayed in red (left panels). North is up. CO 
velocity maps and CO spectra integrated over the features are shown in 
the middle and right panels, respectively. The spectra's units are 
km\,s$^{-1}$ and Jy. Note that the top spectrum 
is integrated only over the $60^\circ$-tilted clump, not the full 
dotted area indicated in the left panel. The CO contour levels are 
$0.12, 0.32, 0.79, 1.9, 5.0$ (top) and $0.11, 0.17, 0.28, 
0.44$~Jy\,beam$^{-1}$\,km\,s$^{-1}$ (bottom). 
}\label{FigLarge}
\end{figure*}

At about 9.5~kpc from the galaxy, we find an extended bright CO 
feature that is tilted $\sim 60^\circ$ from the tail axis (see the 
CO velocity field and integrated CO spectrum in Fig.~\ref{FigLarge}, top 
panel), nearly parallel to the main disk. It is about $10''$ long and 
$3.5''$ wide ($\approx 3.3~{\rm kpc}\times 1.1$~kpc). Its total 
integrated flux is $8\pm 1$~Jy\,km\,s$^{-1}$, corresponding to $\sim 
1.3\times 10^8~M_\odot$. The velocity of the structure ranges from 
$\sim -50$ to 20~km\,s$^{-1}$, with a gradient along its length in the 
N-S direction.

The feature may be part of a longer structure that extends through the 
inner tail and bends upwards (see dotted lines in 
Fig.~\ref{FigLarge}). The fact that it is tilted at a large angle with 
respect to the main tail direction, and is possibly connected to the 
gas in the galaxy, strongly suggests it originated from a large dense 
gas complex in the galaxy, such as a spiral arm, that was gradually 
removed from the disk by ram pressure. The reason is that the flow 
in the downstream direction naturally makes elongated gas features in the 
tail direction, but does not naturally create structures oriented at an 
angle to the tail direction. 
To directly entrain dense, large clumps, dynamical effects can 
help. In a tilted ram pressure orientation, some larger structures 
may have a rotation velocity component in the direction of the wind, 
and in combination with a locally enhanced gravitational potential 
(e.g., by overdensities in spiral arms), they may be pushed gradually 
by the wind instead of being disrupted. 

The $60^\circ$-tilted feature is an example of a dense structure 
of the RPS tail that remained hidden in the previous observations -- 
neither H$\alpha$, nor X-rays show an associated coherent structure, 
and there is also no excess of young stars.
In Sec.~\ref{SecMorphology} we estimated that a gas clump with 
the column density of $30~M_\odot$\,pc$^{-2}$ at 2~kpc disk radius has 
been pushed to $\sim 7$~kpc from the disk in the last 50~Myr. Despite 
the simplicity of the model, it matches well the mean column density 
($\sim 35~M_\odot$\,pc$^{-2}$) and location of the $60^\circ$-tilted 
feature, and thus indicates that the (stripping) age of the feature is 
$\sim 50$~Myr. This time also corresponds to (less than) one half of the 
galaxy rotation period \citep[for the galaxy rotation velocity of $\sim 
110$~km\,s$^{-1}$;][]{jachym2014}. 
The observed shape of the CO structure connecting the 
$60^\circ$-tilted feature to the main galaxy that is not twisted but 
forms a single arc, is consistent with this evolutionary path. 

The main issue for the direct stripping scenario is the survival of 
molecular clouds in the presence of the disruptive effects of the hot ICM wind, notably 
to the Kelvin-Helmholtz (KH) and Rayleigh-Taylor instabilities. The 
crushing time of a cloud with a radius $r_{\rm cloud}$ in a wind with a 
speed $v_{\rm wind}$ can be approximated by the time it takes for a shock 
to propagate through the cloud \citep{klein1994}:
\begin{equation}
t_{\rm cc}= \chi^{1/2}\frac{2\,r_{\rm cloud}}{v_{\rm wind}},
\end{equation}
where $\chi$ is the density contrast between the cloud and the wind. 
For clouds with parameters of typical GMCs the crushing time due to a 
$\sim 1000$~km\,s$^{-1}$ wind is rather short, $\sim 10^7$~yr. For larger clumps 
it is longer, $> 10^8$~yr, which is comparable to their travel (drag) 
time to the inner regions of the tail by ram pressure. Some physical 
processes may protect the clouds and prolonge their lifetime, such as 
magnetic fields which reduce the growth rate of the KH 
instability \citep[e.g.,][]{kauffmann2013}. Also radiative cooling 
effectively extends the lifetime of dense clouds by replenishing the 
cold gas and preventing complete mixing with the hot wind 
\citep[e.g.,][]{cooper2009}.
Since the cooling time is shortest for the densest 
gas, it will be the densest gas clouds that are stripped that will form 
most of the extraplanar molecular clouds, such as the $60^\circ$-tilted 
feature. It is thus 
possible that direct stripping is a viable scenario for the origin 
of the CO-bright large features observed in the inner parts of the tail, 
close to the main galaxy.

\subsection{Large feature parallel to the main tail, with clumpy substructure}
Another large CO feature in the central tail component is a linear 
structure located at nearly 30~kpc from the galaxy, somewhat south of 
the central tail (Fig.~\ref{FigLarge}, bottom panel). It is longer, 
thinner, with more substructure than the $60^\circ$-tilted feature, 
and, most substantially, it is parallel to the main tail 
direction. Along its length of $\sim 
18\arcsec\approx 6$~kpc, at least 8 clumps are identified. Some of the 
clumps are resolved but the width of the feature is mostly unresolved 
at $\sim 1\arcsec$. The total CO flux of nearly 2~Jy\,km\,s$^{-1}$ 
corresponds to $\sim 3.2\times 10^7~M_\odot$. The masses of individual 
clumps are typically $\sim 2.5 - 5\times 10^6~M_\odot$. Their 
characteristic separation in the filament is $\sim 2\arcsec \approx 
0.5-0.6$~kpc. The radial velocity range of the structure is $\sim 0$ to 
$40$~km\,s$^{-1}$ with no apparent velocity gradient along its length 
(although the inner half is at somewhat lower velocity than the outer 
half).

While there is diffuse H$\alpha$ observed along the entire feature, 
there are virtually no young stars or associated compact H$\alpha$ 
peaks. This implies that the CO filament is relatively young, 
unevolved with respect to star formation. Moreover, it is located far 
from the galaxy and is oriented parallel to the main tail. These 
characteristics make it consistent with a dense gas filament formed 
recently in situ in the tail from a less dense gas. It then fragments 
via a gravitational instability, forming a regular substructure of 
molecular clumps, which may eventually form new stars. Thanks to the 
offset location from the main part of the central tail, the CO filament 
is a particularly clear example of a possible in-situ formation.

From the theory of fragmentation of self-gravitating isothermal, 
hydrostatic cylinders, following the analysis in e.g., 
\citet{mattern2018}, we can estimate the most unstable scale: 
\begin{equation}
\lambda_{\rm max}= \frac{22\,c_s}{\sqrt{4\pi G \rho_c}},
\end{equation}
where $\rho_c$ is the central density of a filament in virial 
equilibrium, $c_s= \sqrt{k_B T/\mu m_{\rm H}}$ is the sound speed, and 
$T$, $\mu$, and $m_{\rm H}$ represent the isothermal temperature, the 
mean molecular weight and the mass of hydrogen atom, respectively. For 
the mean density calculated for individual clumps from their masses 
($\sim 2.5\times 10^6~M_\odot$) and sizes ($\sim 350$~pc), and assuming 
$T= 15$~K to calculate the sound speed, the corresponding fragmentation 
lengthscale is $\sim 250$~pc, which is within a factor of $\sim 2-3$ 
consistent with the observed clump separation in the filament. The 
predicted value would be higher and thus match better the observed 
separation when a turbulent medium is assumed, in which the velocity 
dispersion is higher than the sound speed. While these estimates are 
strongly simplified by neglecting the presence of an external pressure, 
rotation, magnetic fields, as well as by using mean densities, they 
suggest that fragmentation due to self-gravity is taking place in the 
stripped gas.

In addition to the central filamentary structure, the most distant CO 
emission region detected at about 58~kpc from the galaxy also has a 
linear morphology. It is shorter, composed of at least three clumps 
with total integrated flux of $\sim 1$~Jy\,km\,s$^{-1}$, thus about 
$1.6\times 10^7~M_\odot$, and is oriented roughly along the tail direction 
(see Fig.~\ref{FigCO}). The flow of matter in the downstream direction 
naturally creates elongated gas features in the tail direction that may 
eventually cool down, condense and fragment. The two CO filamentary 
structures may provide the first direct evidence for molecular gas 
formation in-situ in RPS tails.

The importance of the scenario of molecular gas formation in the tail 
is supported also by the large total amount of molecular gas revealed 
in the tail in the present observations (plus taking into account the 
more extended component; see Sec.~\ref{diffuse}). Following the scaling 
relations of e.g., \citet{catinella2018}, for an unperturbed galaxy 
with a stellar mass $\sim 10^{10}~M_\odot$, the expected total gas 
fraction is $\sim 0.3$, and the molecular-to-atomic gas mass ratio 
$\sim 0.2$, thus the estimated pre-stripping H$_2$ content of 
ESO~137-001 is $\sim 5\times 10^8~M_\odot$. While such estimates may 
vary by almost an order of magnitude at fixed stellar mass, this value 
is well consistent with the detected amount of $\sim 7\times 
10^8~M_\odot$. This suggests that the galaxy is not H$_2$-deficient. 
But it is strongly HI-deficient, following an expected pre-stripping HI 
content of $\sim 2.5\times 10^9~M_\odot$ and a weak upper limit 
obtained from ATCA (the Australia Telescope Compact Array) observations 
of $\sim 5\times 10^8~M_\odot$ \citep{vollmer2001}. Thus, there was in 
principle enough stripped HI to account for the observed H$_2$ content. 

Recent simulations of the 
evolution of cold ($\sim 10^4$~K) gas in a hot galactic wind have shown 
that if the cooling is fast ($t_{\rm cool}\ll t_{\rm cc}$) and if the 
clouds are sufficiently large, they can withstand longer mixing times due 
to the KH instability, and new cold gas may form far downstream from 
the clouds from a ``warm'' mixture of the cold gas and hot wind 
\citep{gronke2018}. Moreover, in the simulations the amount of the 
newly formed cold gas may exceed the original cold gas in the clouds. 
Assuming that the $\sim 10^4$~K gas in the simulations can cool further 
down to reach typical temperatures of molecular gas, this process could 
help to understand the origin of the large total H$_2$ content of 
ESO~137-001.

In order to turn the stripped cold gas (HI) into a molecular phase, the 
presence of dust is crucial. {\it HST} imaging of the galaxy clearly 
shows prominent dust features in the W side of the disk and the inner 
tail. In addition, {\it Herschel} SPIRE imaging (S. Sivanandam et al., 
in preparation) reveals a dust trail emanating from the galaxy in the 
direction of the gas tail. Other optical 
observations (e.g. {\it HST}) of RPS galaxies show clearly dust being 
stripped from the galaxies along with the gas \citep{elmegreen2000, 
abramson2016, cramer2019}. 

Interestingly, the physical scenarios of direct stripping and 
in-situ formation are also relevant for galactic winds, which can have 
a cold gas component associated with the hot gas in outflows \citep[see 
e.g.,][]{banda2019}, and for brightest cluster galaxies, which can have 
molecular gas filaments extending radially from the galaxies 
\citep[e.g.,][]{vantyghem2019}.

\section{Dynamical separation of gas from newly-formed stars in the tail}\label{SecDyn}
Within the tail the degree of correlation between individual gas phases 
and star formation varies (Fig.~\ref{FigHST}). One can find examples of 
CO clumps with and without young stars or HII regions, and vice versa. 
Much of this variety is likely due to different evolutionary stages of 
star formation. While a similar large range in CO/young-star ratios can 
be found in normal galaxies, the distribution of the features in the 
tail and their morphology are particular and characteristic to RPS 
tails. This is caused by dynamical separation -- denser, more compact 
(molecular) clumps are less accelerated by ram pressure and lag behind 
more diffuse gas. Newly formed stars decouple from the surrounding 
gas and if they have not reached the escape velocity, they fall 
back toward the galaxy. Effects of differential acceleration of 
individual phases of the stripped ISM by ram pressure were indicated 
previously from observations \citep[e.g.,][]{vollmer2008, jachym2017}.

Numerous linear streams of young blue stars of different lengths 
parallel with the tail direction which are distributed in all three 
tail components are visible in the {\it HST} image (see 
Fig.~\ref{FigHST}). The ages of 
the young star clusters range from a few Myr to $\sim 100$~Myr and are 
consistent with formation in the tail after stripping. The details 
will be discussed in Waldron, Sun et al. (in prep.). Some stellar and 
gaseous features are physically linked, forming fireballs: a 
star-forming cloud with a linear stream of young stars extending toward 
the galaxy. Fireballs were previously observed in 
several RPS galaxies \citep{cortese2007, yoshida2008, yoshida2012, 
kenney2014}. A simple model of fireballs was introduced by 
\citet{kenney2014}.

\subsection{Fireballs -- linear streams of young stars with molecular clouds at heads}
\begin{figure}
 \centering
 \includegraphics[height=0.198\textwidth]{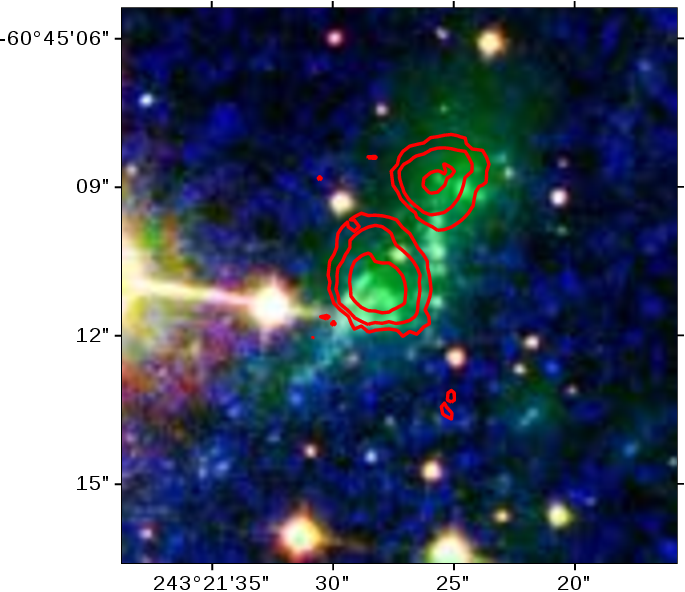}
 \includegraphics[height=0.199\textwidth]{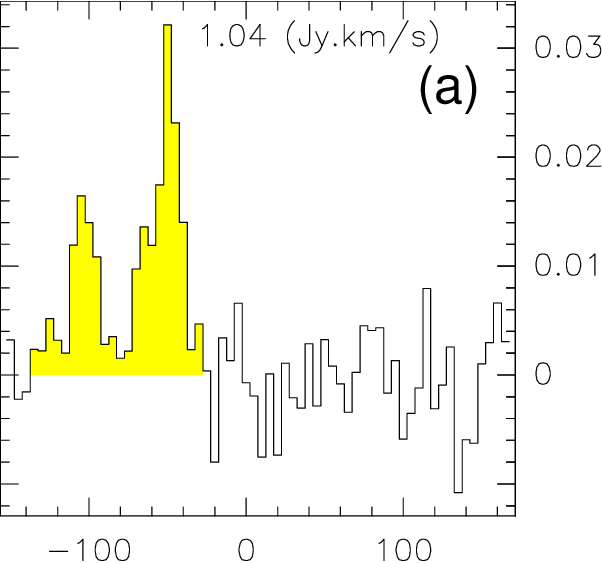}
 \includegraphics[height=0.198\textwidth]{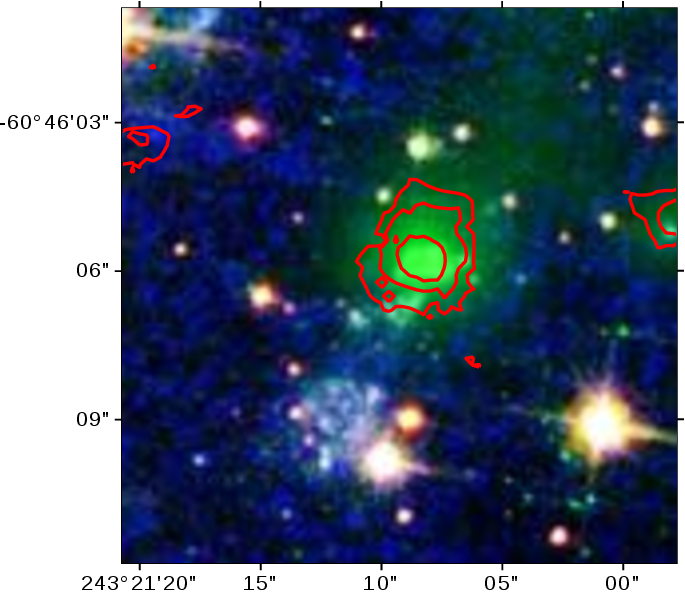}
 \includegraphics[height=0.199\textwidth]{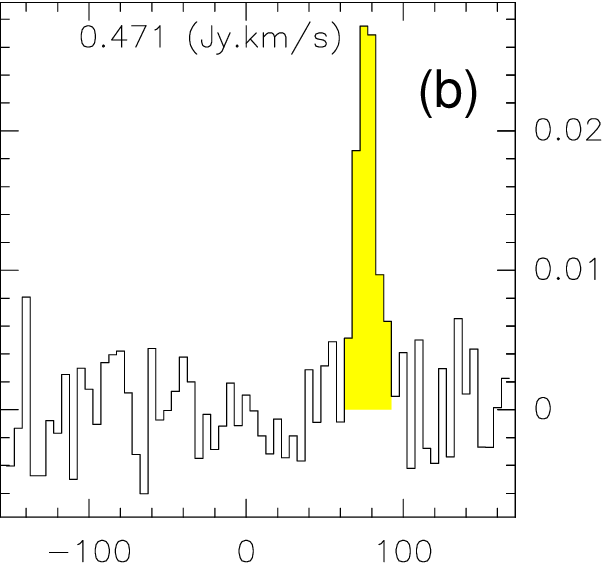}
 \includegraphics[height=0.198\textwidth]{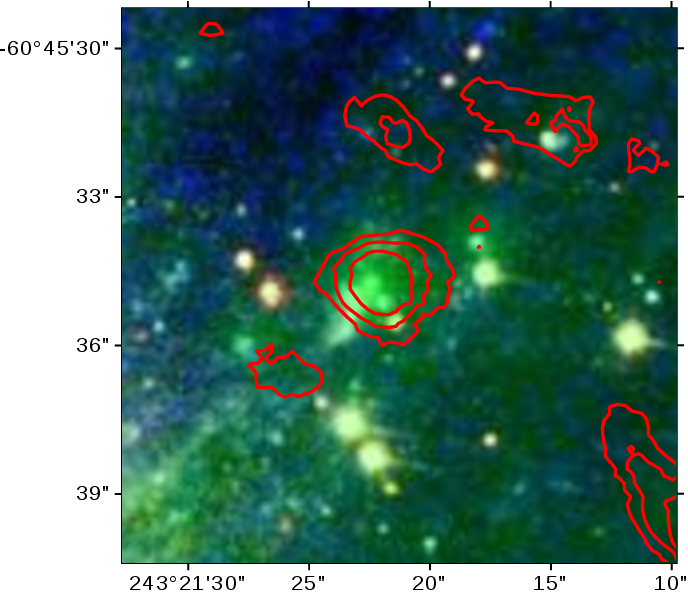}
 \includegraphics[height=0.199\textwidth]{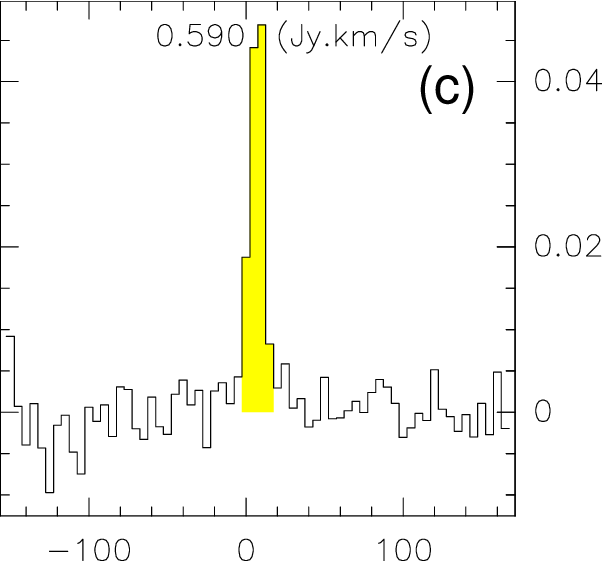}
 \includegraphics[height=0.198\textwidth]{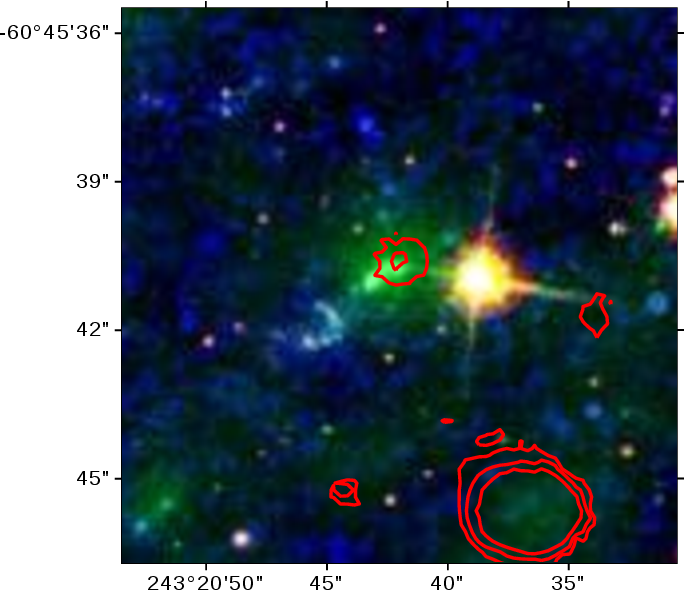}
 \includegraphics[height=0.199\textwidth]{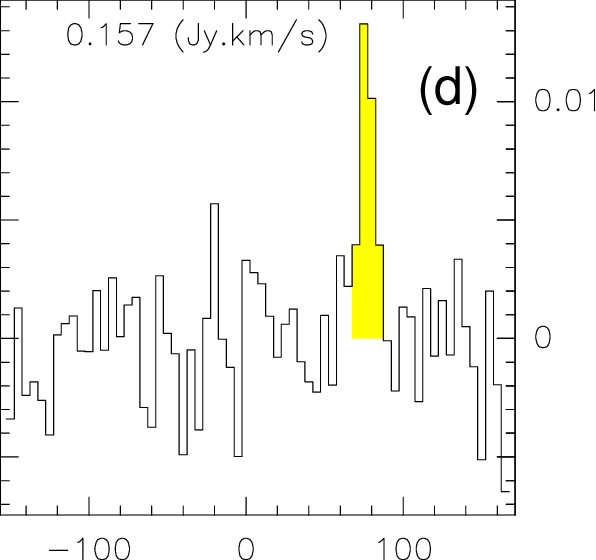}
 \caption{
Examples of small-scale fireballs -- linear streams of young 
stars with molecular clouds at their heads -- from all three tail 
components (see their locations in Fig.~\ref{FigCO} marked with small 
dashed squares). Left panels: CO contours (red) and H$\alpha$ emission 
(green), overlaid on an {\it HST} image (north is up). The CO contour 
levels are $0.11, 0.17, 0.28, 0.44$~Jy\,beam$^{-1}$\,km\,s$^{-1}$. The 
molecular "heads" are associated with compact H$\alpha$ (HII) regions. 
The size of the panels is $\sim 11\times 11$~arcsec$^2$, i.e. $\sim 
3.6\times 3.6$~kpc$^2$. Right panels: CO integrated spectra of the 
central CO features (except the top row where a spectrum integrated 
from both CO clumps is shown). 
}\label{FigFire}
\end{figure}

CO clumps with offset small streams of young stars extending towards 
the galaxy are found in all three tail components. Four examples of 
fireballs are shown in Fig.~\ref{FigFire}. The stellar streams have 
lengths of $\sim 1.5-4.5$~arcsec, i.e., $\sim 0.5-1.5$~kpc and are all 
oriented roughly parallel to the tail direction. The CO clumps at the 
heads of the fireballs are associated with compact H$\alpha$ (HII) 
regions. Some components of fireballs (H$\alpha$ and young stars) have 
previously been observed, but this is the first time the important 
molecular component has been detected.

Fig.~\ref{FigFire} also shows integrated CO spectra of the clumps at 
the heads of fireballs. The integrated CO fluxes are typically $\sim 
0.5$~Jy\,km\,s$^{-1}$, thus about $8\times 10^6~M_\odot$, but some are 
fainter with an integrated CO flux of $\sim 0.16$~Jy\,km\,s$^{-1}$, 
i.e., $2\times 10^6~M_\odot$. The typical sizes of the CO clumps are 
$\sim 2- 2.5''\approx 650- 800$~pc, but some are smaller, $\sim 
500$~pc. Linewidths are $10-15$~km\,s$^{-1}$. Many other fireballs in 
the tail have compact H$\alpha$ clumps at their heads but no detected 
CO emission, similar to fireballs observed previously in other 
galaxies. This may indicate a different evolutionary stage in which 
molecular gas has already been consumed by star formation.

While we cannot measure the velocity structure of the fireballs due to 
the orientation of the orbit of ESO~137-001 in the plane of the sky, it 
is clear that their kinematics evolves with time. It depends on the 
timescales of RP acceleration, gravitational acceleration from the 
galaxy, cloud condensation and star formation. The newly formed stars 
gradually decouple from the gas clump and may eventually start falling 
back toward the galaxy if they have not reached the escape speed. 
Simple modeling with a realistic time-varying ram pressure and a 
galactic potential \citep[adopted from][]{jachym2014} shows that a 
separation of $\sim 1$~kpc between the parent gas cloud and a 
decoupled star (modeled as a high-column-density parcel) establishes 
within typically (a few) 10~Myr. Thus the observed fireballs (including 
those shown in Fig.~\ref{FigFire}) are most likely dynamically young 
structures with molecular clumps at their head that have been actively 
forming new stars over the past $\sim 10$~Myr. 

\begin{figure}
\centering
 \includegraphics[width=0.45\textwidth]{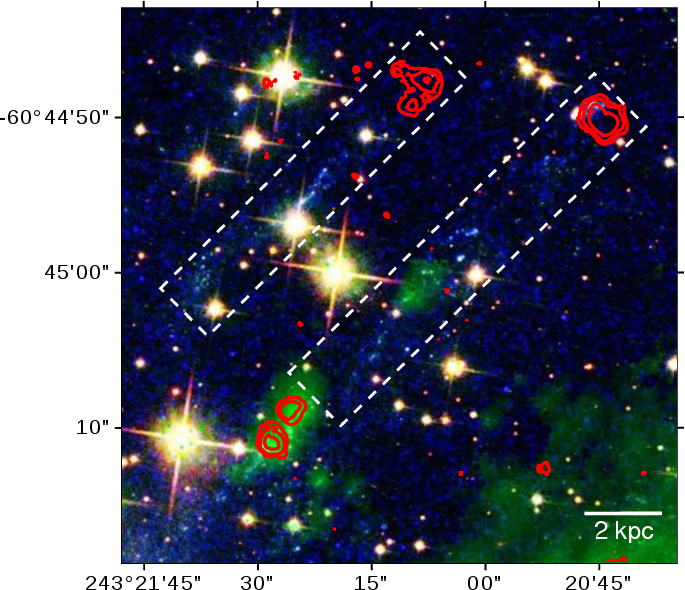}
\caption{
Large-scale fireballs (superfireballs) in the northern component of the 
tail. CO contours (red; $0.12, 0.20, 
0.32$~Jy\,beam$^{-1}$\,km\,s$^{-1}$) are overlaid on H$\alpha$ (green) 
and a background {\it HST} image.
}\label{FigSuper}
\end{figure}

\subsection{Superfireballs}
In the northern component of the tail, close to the disk, we find possible 
examples of much larger fireballs. In this part of the tail, which 
likely is in a later evolutionary stage of stripping, there is no 
diffuse X-ray or H$\alpha$ emission, but distinct streams of young 
stars, HII regions (compact H$\alpha$ sources), and bright CO clouds. 
Most of the young stars seen in the {\it HST} image in this region fall 
within two linear streams $\sim 15-25''\approx 5-8$~kpc in length, 
parallel to the main tail (see a closeup in Fig.~\ref{FigSuper}).

Each of these streams have CO clouds coincident with or somewhat beyond 
the outermost stars (in the downstream direction). This is consistent 
with the fireball scenario. The CO clouds have diameters of $\sim 
2\arcsec$ and $4\arcsec$, i.e., $\sim 650$~pc and $1300$~pc (but with 
possible substructure); their linewidths are $\sim 22$ and 
$30$~km\,s$^{-1}$, and H$_2$ masses $\sim 6.4\times 10^6$ and 
$1.9\times 10^7~M_\odot$, respectively for the E and W structures. The 
long streams do not have young stars extending continuously to the CO 
cloud or along the length of the streams. 
This indicates that molecular clouds in tails may survive after a 
period of active star formation ends, and possibly go through several 
more phases of star formation activity. For example, the western 
fireball shows an H$\alpha$ compact feature (with no 
detected associated CO emission) in its inner half. If not simply due 
to the effects of projection, this could correspond to a previous 
episode of condensation when denser (pre-stellar) fragments decoupled 
from a parental clump. The latter continued to be accelerated further 
by ram pressure in the downstream direction and condensed again only 
recently. The multiple features are then parts of one large-scale 
structure. In this picture small-scale fireballs in fact can be part 
of large-scale fireballs. We note that the locations of the two 
CO clumps at the heads of the large fireballs lay outside the MUSE area 
but previous SOAR observations show bright compact H$\alpha$ emission 
associated with at least the western CO clump \citep{sun2007}. 

Superfireballs, due to their lengths, are expected to be dynamically 
older structures than the small-scale fireballs. This is consistent 
with their location in the northern outer tail component. Our modeling 
shows that $\sim 50$~Myr is needed to develop a separation of $\sim 
10$~kpc between the parent cloud and decoupled stars. 
We may also be seeing superfireballs in the process of formation in the 
southern tail component. There the sensitive MUSE H$\alpha$ observations 
reveal many elongated structures (see Fig.~\ref{FigHST}), some of which 
show a head-tail morphology -- with diffuse tails extening away from 
the galaxy and compact clumps at the heads situated closer to the 
galaxy (sometimes associated with CO emission). 
Some are aligned into long coherent 
structures, encompassing other H$\alpha$ compact clumps. It is possible 
that with ongoing ram pressure the diffuse component will be cleared 
out of the structures and moved downstream along the tail, leaving 
behind 'nude' structures of denser clumps and stellar streams, i.e., 
superfireballs.

\subsection{Double-sided fireballs}
\begin{figure}
\centering
 \includegraphics[height=0.45\textwidth]{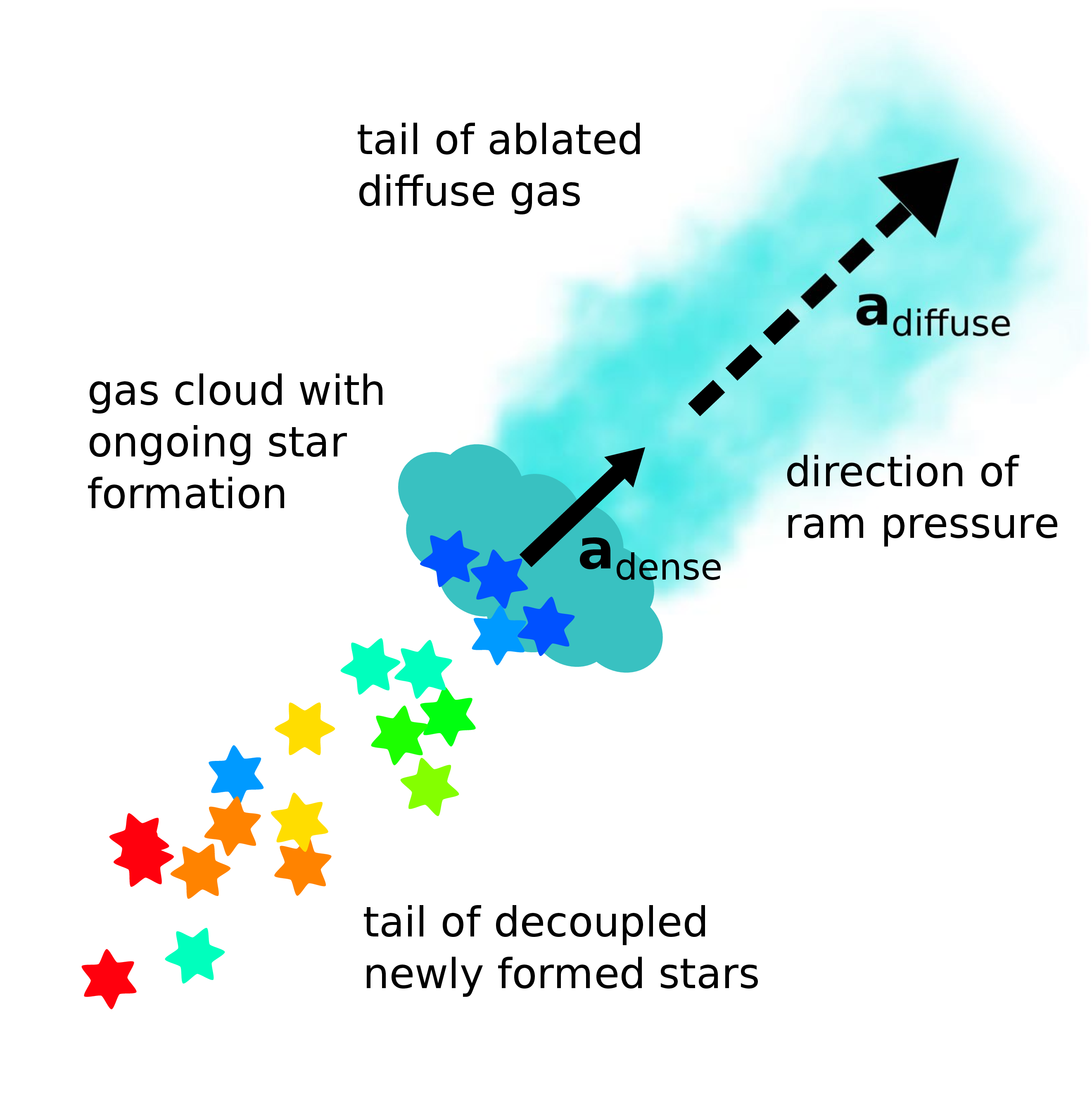}
\caption{
A schematic of the double-sided fireball model -- with a tail of 
newly-born stars pointing toward the galaxy, decoupled from the 
outwardly accelerating gas cloud, and another tail of diffuse gas 
ablated by ram pressure from the denser cloud extending in the opposite 
direction. The diffuse gas with a lower surface density is accelerated 
more than the dense gas (as indicated in the picture by the arrows). In 
this simple scenario an age gradient would form along the stellar tail, 
but further fragmentation and cooling/collapse of the decoupled 
star-forming dense clumps may complicate it.
}\label{FigDouble}
\end{figure}

Some of the H$\alpha$ head-tail features in the tail are identified to 
be interconnected with fireballs. In addition to the \citet{kenney2014} 
model, the tail of ESO~137-001 indicates that the fireballs may be 
double-tail features, with a second, gaseous tail of diffuse 
(H$\alpha$) gas that is ablated from the dense star-forming clump that 
extends in the opposite direction to the stellar tail - away from the 
galaxy. The central dense clump of a fireball is bright in H$\alpha$ 
and possibly also in CO emission. A cartoon of the model of the 
double-sided fireballs is shown Fig.~\ref{FigDouble}. Similar features 
of considerable lengths were previously found in the RPS tail of the 
Coma galaxy RB199 \citep{yoshida2008}, although without any associated 
molecular component yet observed.

\begin{figure}
 \centering
 \includegraphics[width=0.45\textwidth]{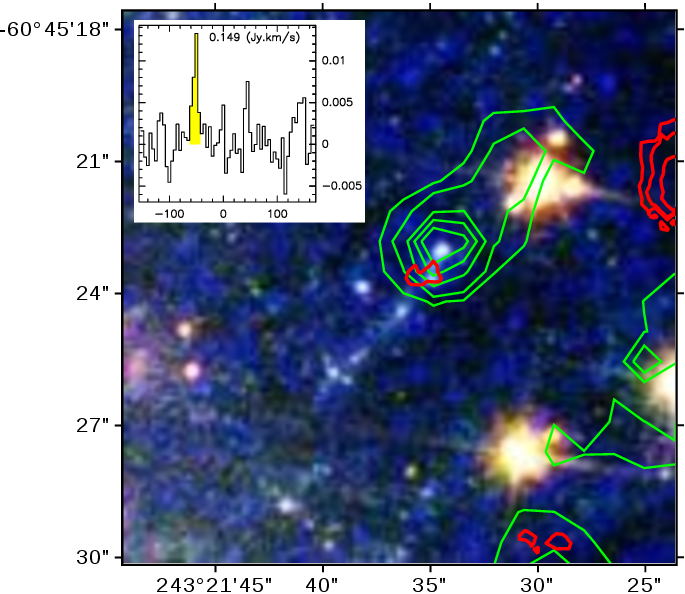}
 \caption{
Example of a double-sided fireball in a {\it HST} image with H$\alpha$ 
contours (green) and CO contours (red) overlaid. The 
feature consists of a compact HII region associated with weak CO 
emission, a stellar tail pointing toward the galaxy and a tail of more 
diffuse H$\alpha$ emission extending in the opposite direction. The CO 
contour levels are $0.1, 0.15, 0.24, 
0.39$~Jy\,beam$^{-1}$\,km\,s$^{-1}$. The displayed region is $\sim 
12.5\arcsec\times 12.5\arcsec$. The inset panel shows the spectrum 
of the central associated CO emission region.
}\label{FigDoubleObs}
\end{figure}

An example of a double-sided fireball is given in 
Fig.~\ref{FigDoubleObs} which shows a compact HII region associated 
with a (slightly offset) weak CO clump at the head of a small fireball 
with a diffuse tail extending along the direction of the wind. It is 
located in the northern component of the tail, close to the main 
galaxy. The associated CO emission region has an H$_2$ mass of $\sim 
2.4\times 10^6~M_\odot$. More similar structures (though less clear and 
without detected associated CO clumps) may be found mainly in the 
southern tail component.

\section{Conclusions}\label{SecCon}
We observed with ALMA in CO(2-1) the ram pressure stripped tail of 
the Norma cluster galaxy ESO~137-001, one of the nearest jellyfish
galaxies. The observations covered at $\sim 1\arcsec\approx 350$~pc 
resolution nearly the full extent of the multi-component H$\alpha$ and 
X-ray tail - the central tail, the northern 'orphaned' regions, as well 
as the southern tail component. It is the first time that a 
high-resolution map of cold molecular gas in a ram pressure stripped 
tail has been acquired and together with observations at other 
wavelengths it offers the most complete view of the spectacular RPS 
tail to date. The main results of a first analysis of the CO map and 
a comparison with observations at other wavelengths are:

CO emission arises from a rich distribution of mostly compact clumps 
detected in the tail out to nearly 60~kpc distance from the disk, and 
over a lateral extent of about 24~kpc. The CO clumps are associated 
with all three components of the tail that are likely in different 
evolutionary stripping stages. 

Most of the detected CO clumps in the tail are larger structures with 
masses of Galactic GMCs and GMAs, but with sizes and linewidths that 
are larger than is typical of those found in the Galaxy. Estimates 
of their virial parameters indicate that 
the molecular clumps are not gravitationally bound and will disperse 
with time. The CO velocity field in the tail is dominated by the 
galaxy's rotation that is imprinted into the stripped gas -- most of 
the northern regions are blue-shifted, while those in the southern tail 
component are mostly red-shifted, relative to the tail centroid.

In total $\sim 9\times 10^8~M_\odot$ of H$_2$ was revealed in the tail, 
assuming the standard Galactic CO-to-H$_2$ conversion factor. 
Comparison of fluxes of the ALMA+ACA observations with previous 
single-dish (APEX) observations indicate that in addition to the 
compact CO features there is a substantial component of extended 
(scales $>18\arcsec$) molecular gas in the tail. Its 
fraction increases with distance from the parent galaxy. 
ESO~137-001 is molecular rich -- the total amount of H$_2$ detected 
in the galaxy and the tail likely exceeds the pre-stripping molecular 
content of the galaxy.

From the morphology of some CO features we can infer their origin in 
the tail. For example, filamentary structures oriented parallel to the 
tail direction and located at large distances from the disk have likely 
formed in-situ, through fragmentation via 
gravitational instability. Other CO features, such as the large clump 
in the inner central tail, tilted $\sim 60^\circ$ from the tail axis, 
may have formed from dense gas clumps, such as a spiral arm, stripped 
gradually from the disk. These CO features do not form distinct 
structures in H$\alpha$ or X-ray emisison, and there is also no 
associated excess of young star formation.
Simple analysis estimates their ages to $\sim 50$~Myr.

The overall CO distribution in the tail roughly follows the 
distribution of the other gas phases and the regions of young star 
formation, but at small scales the degree of correlation varies. It 
is the distribution of the features and their morphology that are 
particular and characteristic to RPS tails. Numerous 'fireball' 
features with CO clumps at their heads and tails of young stars 
(typically $\sim 1.5$~kpc) pointing towards the main galaxy are found 
in all three tail components. 
Newly formed stars decouple from the 
surrounding gas -- those in the outer tail probably escape from the 
galaxy but the inner tail stars fall back onto the galaxy leading to halo 
stellar streams. Several much longer fireballs (lengths of $5-8$~kpc) 
are also observed, especially in the northern regions of the tail, that 
may be in an older evolutionary stripping phase. The small fireballs 
are likely dynamically young structures ($\sim 10$~Myr) while 
their larger relatives have ages of $\sim 50$~Myr.

We found that some linear H$\alpha$ features, compact H$\alpha$ 
emitting clumps, and young star regions may be interconnected into 
double-sided fireballs. We propose a simple scenario based on 
dynamical separation of gas phases 
in which denser gas clumps pushed by ram pressure form stars that 
decouple from the clumps and create inward-pointing stellar tails, 
while at the same time the clumps are ablated by the strong ram 
pressure and tails of diffuse gas are formed in the opposite direction 
to the stellar tails.

\acknowledgments
This paper makes use of the following ALMA data:
ADS/JAO.ALMA\#2015.1.01520.S. ALMA is a partnership of ESO 
(representing
its member states), NSF (USA) and NINS (Japan), together with NRC
(Canada), MOST and ASIAA (Taiwan), and KASI (Republic of Korea), in
cooperation with the Republic of Chile. The Joint ALMA Observatory is
operated by ESO, AUI/NRAO and NAOJ. 
This research has made use of the HyperLeda database
(\url{http://leda.univ-lyon1.fr}).

We thank our anonymous referee for helpful comments that improved the 
quality of this paper.
P.J. acknowledges support through project LM2015067 of the Ministry of
Education, Youth and Sports of the Czech Republic, and the
institutional research project RVO:67985815. 
M.S. acknowledges support from NASA grants GO6-17111X and 80NSSC18K0606 
and NSF grant 1714764. 
L.C. is the recipient of an Australian Research Council Future 
Fellowship (FT180100066) funded by the Australian Government. 
T.S. acknowledges the support through fellowship SFRH/BPD/103385/2014 
funded by FCT (Portugal) and POPH/FSE (EC). This work was supported by 
Funda\c{c}\~{a}o para a Ci\^{e}ncia e a Tecnologia (FCT) through 
national funds (UID/FIS/04434/2013) and by FEDER through COMPETE2020 
(POCI-01-0145-FEDER-007672). 
T.S. also acknowledges support from DL 57/2016/CP1364/CT0009. This work 
was supported by FCT/MCTES through national funds (PIDDAC) by this 
grant PTDC/FIS-AST/29245/2017.
E.B. acknowledges support from the UK Science and Technology Facilities 
Council [grant number ST/M001008/1].
M.F. acknowledges support by the Science and Technology Facilities 
Council [grant number ST/P000541/1]. This project has received funding 
from the European Research Council (ERC) under the European Union's 
Horizon 2020 research and innovation programme (grant agreement No 
757535).


\vspace{5mm}
\facility{ALMA}


\software{CASA \citep{mcmullin2007}, GILDAS \citep{guilloteau2000}, 
astropy \citep{astropy2013}}

\end{document}